\documentclass[10pt,letterpaper]{article}
\usepackage{opex3,color,upgreek,amsmath,amssymb,cite}

\def\degree{\mbox{$^\circ$}}

\def\ii{{\mathrm{i}}}

\def\ee{{\mathrm{e}}}
\def\dd{{\mathrm{d}}}

\def\no{{\nonumber}} 

\def\sub#1{_\mathrm{#1}} 

\DeclareGraphicsExtensions{.eps}

\begin{document}

\title{Classical realization of dispersion-canceled, \\
artifact-free, and background-free \\
optical coherence tomography}

\author{Kazuhisa Ogawa$^{1,2,*}$ and Masao Kitano$^{1}$}

\address{$^1$Graduate School of Engineering, Kyoto University,\\
Kyoto daigaku-katsura, Nishikyo-ku, Kyoto 615-8510, Japan\\
$^2$Graduate School of Information Science and Technology, Hokkaido University,\\ Kita 14, Nishi 9, Kita-ku, Sapporo 060-0814, Japan}

\email{$^*$ogawak@ist.hokudai.ac.jp} 



\begin{abstract}
Quantum-optical coherence tomography (Q-OCT) provides a dispersion-canceled axial-imaging method, but its practical use is limited by the weakness of the light source and by artifacts in the images. 
A recent study using chirped-pulse interferometry (CPI) has demonstrated dispersion-canceled and artifact-free OCT with a classical system; however, unwanted background signals still remain after removing the artifacts.  
Here, we propose a classical optical method that realizes dispersion-canceled, artifact-free, and background-free OCT.
We employ a time-reversed system for Q-OCT with transform-limited input laser pulses to achieve dispersion-canceled OCT with a classical system.
We have also introduced a subtraction method to remove artifacts and background signals.
With these methods, we experimentally demonstrated dispersion-canceled, artifact-free, and background-free axial imaging of a coverglass and cross-sectional imaging of the surface of a coin.
%
\end{abstract}



\bibliographystyle{osajnl}





\section{Introduction}

Optical coherence tomography (OCT) \cite{huang1991optical} is an axial-imaging technique that uses white-light interference.
OCT has been widely used for various imaging applications in the medical \cite{fujimoto1995optical,hee1995optical,brezinski1997assessing,welzel2001optical,vakoc2012cancer,fercher2003optical,podoleanu2012optical} and industrial fields \cite{dufour2005low,walecki2005novel,webster2007high}.
The axial resolution of OCT is ultimately limited by the coherence length of the light source; however, dispersion broadens the signal width and results in degradation of the axial resolution \cite{drexler2001ultrahigh}. 
Various techniques that have been used for depth-dependent dispersion compensation \cite{doi:10.1117/1.429900,smith2002real,fercher2001numerical,fercher2002dispersion} require \textit{a priori} knowledge of the dispersion of the sample.

Quantum-optical coherence tomography (Q-OCT) \cite{PhysRevA.65.053817,PhysRevLett.91.083601,nasr2009quantum} is a way to avoid resolution degradation by dispersion without \textit{a priori} knowledge of the sample properties.
Q-OCT employs Hong--Ou--Mandel (HOM) interference \cite{PhysRevLett.59.2044} of time-frequency-entangled photon pairs; its interferogram (HOM dip) is insensitive to even-order dispersion such as group-velocity dispersion (GVD) \cite{PhysRevA.45.6659,PhysRevLett.68.2421}.
Using Q-OCT, dispersion-canceled axial imaging have been demonstrated for a coverglass \cite{PhysRevLett.91.083601} and the surface topography of a gold-coated onionskin \cite{nasr2009quantum}.
However, the practical application of Q-OCT is limited by two obstacles: the weakness of the output signals and the inclusion of artifacts.
The weakness of the output signals stems from the low efficiency in the generation of entangled photon pairs by spontaneous parametric down-conversion (SPDC) and the necessity of coincidence counting of these photon pairs. 
Weak photon flux requires long-term measurements, which prevents Q-OCT from imaging of living samples.
The artifacts are observed between each pair of HOM dips, corresponding to interfaces of the sample (in this paper these proper HOM dips are called \textit{main signals} in order to distinguish them from artifacts).
Because the number of artifacts is roughly equal to the square of the number of interfaces, the Q-OCT images of complex samples are cluttered by many artifacts.

Recent studies have shown that dispersion-canceled HOM-like dips can be reproduced by a time-reversed HOM interferometer using pairs of oppositely chirped laser pulses (chirped-pulse interferometry, CPI \cite{kaltenbaek2008quantum}) and by that using pairs of transform-limited laser pulses with various time differences (time-resolved pulse interferometry, TRPI, in our previous work \cite{PhysRevA.91.013846}). 
Both techniques are implemented by classical-optical systems and therefore enable us to use an intense light source.
CPI, furthermore, has been applied to dispersion-canceled OCT (CPI-based OCT), such as axial imaging of a coverglass \cite{Lavoie2009} and cross-sectional imaging of an onion piece \cite{mazurek2013dispersion}; the latter experiment was also able to remove artifacts.   
Other classical techniques have been reported, showing automatic dispersion cancellation \cite{resch2007classial,banaszek2007blind,PhysRevA.74.041601,le2010experimental,shirai2014intensity,ryczkowski2015experimental}, but so far none of them have reached practical axial imaging and they still suffer from artifacts.  

However, even in the artifact-free CPI-based OCT experiments \cite{mazurek2013dispersion}, there tend to remain \textit{background signals} around the main signals in the OCT images.
Even though the background signals are relatively low compared to the main signals, the background signals lead to blurred OCT images, especially in log-scale representations. (The log-scale representations are usually used to exhibit practical OCT images to enhance the main signals \cite{huang1991optical,fujimoto1995optical,hee1995optical,brezinski1997assessing,welzel2001optical,vakoc2012cancer,fercher2003optical,podoleanu2012optical,dufour2005low,walecki2005novel,webster2007high}.)
Therefore, the artifact-free CPI-based OCT method needs another procedure to remove the background signals. 

In this paper, we propose a classical-optical OCT method that can achieve dispersion-canceled, artifact-free, and background-free OCT images.
We employ an OCT method using TRPI (TRPI-based OCT), which can produce intense and dispersion-canceled OCT images like the CPI-based OCT.
In addition, we introduce a technique called \textit{subtraction method} to remove the artifacts and the background signals around main signals.
Combining TRPI-based OCT and the subtraction method, we experimentally demonstrate dispersion-canceled, artifact-free, and background-free axial imaging of a coverglass and cross-sectional imaging of the surface of a coin.
We also discuss the advantages and drawbacks of TRPI-based OCT with the subtraction method compared with CPI-based OCT.

\section{Methods and characterization}\label{sec:theory6}

In this section, we describe TRPI-based OCT with the subtraction method realizing dispersion-canceled, artifact-free, and background-free axial imaging.
We also provide experimental results for axial imaging of a coverglass to characterize these methods.
In Sec.~\ref{sec:disp-canc-oct}, we introduce a TRPI-based OCT system that simply uses HOM-like dips of the time-reversed HOM interferometer \cite{PhysRevA.91.013846} as main signals, and note that this OCT system suffers from artifacts.
In Sec.~\ref{sec:removal-artifacts}, we describe the subtraction method to remove the artifacts and the background signals.

\subsection{TRPI-based OCT}\label{sec:disp-canc-oct}

\begin{figure}
\begin{center}
\includegraphics[width=31.5pc]{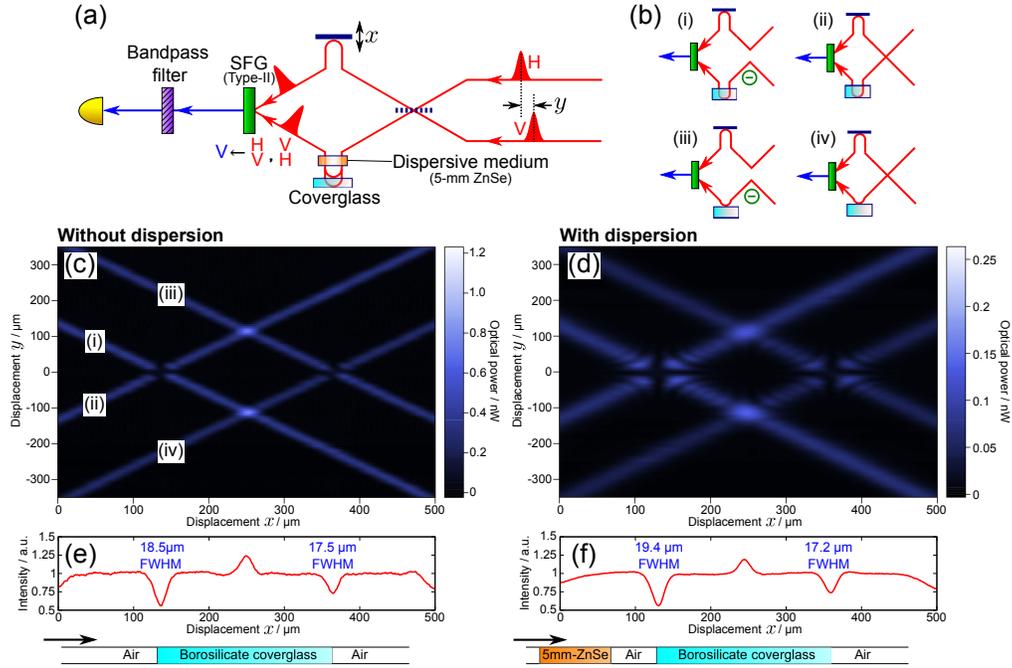}
\end{center}
\caption{
(a) Schematic setup of the TRPI-based OCT system using the time-reversed HOM interferometer. 
(b) Feynman paths leading to output signals in this setup.
(c)--(f) Measurement results without and with the dispersive medium.
(c), (d) Measured intensity distributions $I(x,y)$.  
The four bright lines (i)--(iv) correspond to the four Feynman paths (i)--(iv) in (b).
(e), (f) Interferograms derived by integrating the intensity distributions $I(x,y)$ with respect to $y$. This shows the axial images of the coverglass.
Each interferogram is dispersion-insensitive, but has an artifact at the center of the two main signals. 
}\label{fig:6-3}
\end{figure}

We first consider a TRPI-based OCT system that simply uses HOM-like dips of the time-reversed HOM interferometer \cite{PhysRevA.91.013846} as main signals.
The schematic is shown in Fig.~\ref{fig:6-3}(a).
This system has the same composition as the time-reversed HOM interferometer \cite{PhysRevA.91.013846}, except that the mirror in the lower arm is replaced with a measured sample.
The input light is two orthogonally polarized pulses separated by a variable distance $y$.
The two pulses enter the cross-correlator; the reference arm includes a delay line to introduce a relative path difference $x$, and the sample arm includes a sample and a dispersive medium. 
The pulses are converted into sum-frequency light by type-II sum-frequency generation (SFG).
The sum-frequency light is fed into a narrow bandpass filter and detected by a photodiode.
We measure the optical intensity $I(x,y)$ for various $x$ and $y$, and integrate it with respect to $y$.
Then we obtain a dispersion-canceled OCT profile, in which the HOM dips are observed at the positions corresponding to the interfaces of the sample.

We provide a brief theoretical analysis of this system.
We describe the complex field amplitudes of the input two pulses in the upper and lower input ports as $E(\omega)=\exp[-(\omega-\omega_0)^2/(2\Delta\omega^2)]$ and $E(\omega)\ee^{\ii\omega y/c}$, respectively, where $\omega_0$, $\Delta\omega$, and $c$ are the central frequency, the root-mean-square (RMS) width of the spectrum, and the speed of light in a vacuum, respectively. 
After the beam splitter, the field amplitudes receive a phase shift of $\ee^{\ii\omega x/c}$ in the reference arm, and a linear transfer function $H(\omega)$ in the sample arm.
$H(\omega)$ models the effect of both the sample with multiple interfaces and the dispersive medium.
The field amplitudes are converted, by type-II SFG, into the following convolution integral:
\begin{align}
E\sub{SFG}(\omega)
\propto&
\int^{\infty}_{-\infty}\dd\omega'
E(\omega')H(\omega')\cdot E(\omega-\omega')\ee^{\ii(\omega-\omega')(x+y)/c}\no\\
&\hspace{2cm}-\int^{\infty}_{-\infty}\dd\omega'E(\omega')H(\omega')\ee^{\ii\omega'y/c}\cdot E(\omega-\omega')\ee^{\ii(\omega-\omega')x/c}.
\end{align}
Assuming that the transmission spectrum of the bandpass filter is sufficiently narrow and its center frequency is $2\omega_0$, the measured intensity $I(x,y)$ after the bandpass filter is given by $I(x,y)\propto|E\sub{SFG}(2\omega_0)|^2$. 
By integrating $I(x,y)$ with respect to $y$, the post-processed interferogram is calculated as 
\begin{align}
S(x):=&\int^{\infty}_{-\infty}\dd yI(x,y)\no\\ 
\propto&
\int^{\infty}_{-\infty}\dd\omega'
|E(\omega_0+\omega')|^2|E(\omega_0-\omega')|^2|H(\omega_0+\omega')|^2\no\\
&\hspace{1cm}-\int^{\infty}_{-\infty}\dd\omega'
|E(\omega_0+\omega')|^2|E(\omega_0-\omega')|^2H(\omega_0+\omega')H(\omega_0-\omega')^*\ee^{-\ii 2\omega'x/c}.\label{eq:1}
\end{align}
We now assume that the sample is a coverglass behind a dispersive medium.
The linear transfer function is given by $H(\omega)=\ee^{\ii\phi(\omega)}\left(r_1+r_2\ee^{\ii\omega2dn/c}\right)$, where $r_1$ and $r_2$ are the reflectances of the top and bottom surfaces, respectively. In addition, $d$ and $n$ are the thickness and refractive index of the coverglass, respectively. 
We can see that the effect of the even-order dispersion $\phi(\omega)=\phi_0+\phi_2(\omega-\omega_0)^2+\cdots$ is canceled out and $S(x)$ is explicitly calculated as
\begin{align}
S(x)
\propto\,&
r_1^2+r_2^2+2r_1r_2\cos\frac{\omega_02dn}{c}\exp\frac{-(dn)^2}{2(c/\Delta\omega)^2}
\no\\
&\hspace{0.1cm}
-r_1^2\exp\frac{-x^2}{2(c/\Delta\omega)^2}
-r_2^2\exp\frac{-(x-2dn)^2}{2(c/\Delta\omega)^2}
-2r_1r_2\cos\frac{\omega_02dn}{c}\exp\frac{-(x-dn)^2}{2(c/\Delta\omega)^2},
\end{align}
where the first to third terms are constants, the fourth and fifth terms are HOM-like dips (main signals) corresponding to the two surfaces of the coverglass, and the sixth term is an artifact.
The artifacts appear as dip- or peak-shapes, depending on $\omega_0$, $d$, and $n$. The dip-shape artifact is indistinguishable from the main signals.

We experimentally performed an axial imaging of a coverglass with this TRPI-based OCT system to evaluate the performance of this system.
In the experimental setup shown in Fig.~\ref{fig:6-3}(a), we used a femtosecond fiber laser (center wavelength 782\,nm, pulse duration 74.5\,fs FWHM, average power 54\,mW, repetition rate 100\,MHz) and a 5-mm-thick ZnSe plate (GVD$=$1075\,fs$^2$/mm at 782\,nm) as a dispersive medium. In addition, we used a 1-mm-length $\upbeta$-barium borate (BBO) crystal for type-II non-collinear SFG, a 1200-lines/mm aluminum-coated diffraction grating followed by a slit as a bandpass filter (0.40-nm bandwidth around 391\,nm), and a Si photodiode (see Fig.~\ref{fig:6-6} in Sec.~\ref{sec:experiments-results6} for the detailed experimental configuration).
Figures~\ref{fig:6-3}(c)--(f) show the measurement results without and with the dispersive medium. 
Figures~\ref{fig:6-3}(c) and (d) are the measured intensity distributions $I(x,y)$; they are obtained by scanning the delay $x$ for each value of $y$ changed by a step of 5\,$\upmu$m. 
The optical power is in the order of sub-nanowatt, which corresponds to about $10^9$\,photons/s and is three orders of magnitude greater than that in the previous Q-OCT experiments \cite{nasr2009quantum}. 
In each intensity distribution, we can see four bright lines, which correspond to the four Feynman paths (i)--(iv) shown in Fig.~\ref{fig:6-3}(b).
The pairs of paths [(i), (ii)] and [(iii), (iv)] have a phase difference $\pi$ and interfere destructively.
On the other hand, the pairs of paths [(i), (iv)] and [(ii), (iii)] have the phase difference $\pi+\omega_02dn/c$, and in this case, interfere somewhat constructively.
Figures~\ref{fig:6-3}(e) and (f) are axial images of the coverglass, which are derived by integrating $I(x,y)$ with respect to $y$. 
Each interferogram has two HOM-like dips of main signals and a peak-shaped artifact at the center of the two dips, as is the case in Q-OCT \cite{PhysRevLett.91.083601} and CPI-based OCT \cite{Lavoie2009}. 
Even with dispersion, the interferogram in Fig.~\ref{fig:6-3}(f) remains essentially unchanged due to automatic dispersion cancellation, whereas the four lines (i)--(iv) in Fig.~\ref{fig:6-3}(d) are broadened.
As seen in the intensity distributions of Figs.~\ref{fig:6-3}(c)--(f), the artifacts are attributed to the interference between the lines (i) and (iv) and between (ii) and (iii).
Therefore, we can remove the artifacts by avoiding these intersections when integrating the intensity distributions; the detail of this technique will be introduced in the next section.

\subsection{Subtraction method}\label{sec:removal-artifacts}

We introduce the subtraction method to make TRPI-based OCT free from artifacts and from background signals remaining around the main signals after removing the artifacts.
By combining TRPI-based OCT described in Sec.~\ref{sec:disp-canc-oct} and the subtraction method, we experimentally perform an axial imaging of a coverglass and demonstrate not only dispersion-canceled, but also artifact- and background-free OCT.
We used the same experimental apparatus as in the previous section.

\begin{figure}
\begin{center}
\includegraphics[width=31.5pc]{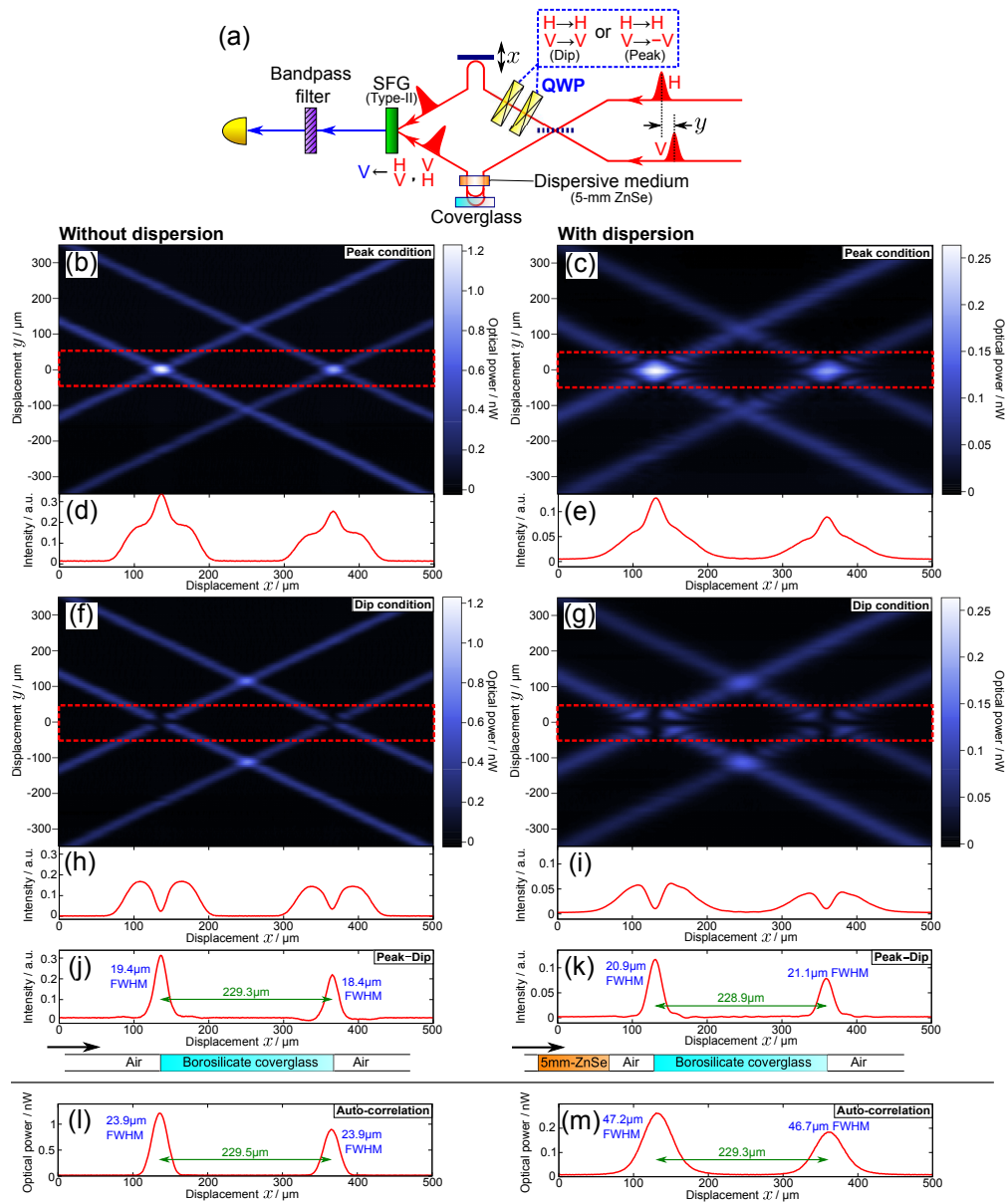}
\end{center}
\caption{
(a) Schematic setup implementing the subtraction method.
One of the QWPs can be rotated by 90$\degree$ to switch the peak and dip conditions.
(b)--(e), (f)--(i) Experimental results in the peak and dip condition, respectively, without and with the dispersive medium.
(b), (c), (f), (g) Measured intensity distributions $I(x,y)$.
(d), (e), (h), (i) Interferograms derived by integrating the intensity distributions from $y=-50\,\upmu$m to $50\,\upmu$m.
The integrated domain is indicated by the area enclosed by a red dashed line in (b), (c), (f), and (g).  
(j), (k) Interferograms derived by subtracting (h) and (i) from (d) and (e), respectively, which shows dispersion-canceled, artifact-free, and background-free main signals.
(l), (m) Interferograms of the auto-correlation, derived from the intensity distributions in the peak condition (b) and (c) at $y=0\,\upmu$m.
}\label{fig:6-4}
\end{figure}

We describe the procedure of the subtraction method with the optical system shown in Fig.~\ref{fig:6-4}(a).
This system is same as the TRPI-based OCT system described in Sec.~\ref{sec:disp-canc-oct}, except for two quarter-wave plates (QWPs) inserted in the reference arm.
When the two QWPs' fast axes are parallel, the pair of QWPs acts as a half-wave plate (HWP) and multiplies the vertically-polarized light by $-1$.
On the other hand, when the two QWPs' fast axes are orthogonal, the effects of the two QWPs cancel each other.
We call these conditions \textit{peak} and \textit{dip conditions}, respectively.
We measure the optical intensity $I(x,y)$ for various $x$ and $y$ and for both conditions.

The intensity distributions $I(x,y)$ in the peak condition are shown in Figs.~\ref{fig:6-4}(b) and (c).
Compared with those in Figs.~\ref{fig:6-3}(c) and (d), the constructive and destructive interferences switch places due to the QWPs.
We integrate each intensity distribution over a limited range of $y$ around zero (from $y=-50\,\upmu$m to $50\,\upmu$m in this case) to avoid that the intersection areas lead to generation of artifacts.
The derived interferograms are shown in Figs.~\ref{fig:6-4}(d) and (e).
We can see that the main signals are peak-shaped and insensitive to dispersion and that the artifacts are removed, but the background signals remain around the main signals.
We note that this integration range of $y$ must cover most part of the intersection areas leading the main signals.
In the presence of dispersion, the intersection areas are broadened, as seen in Fig.~\ref{fig:6-4}(c), and the integration range of $y$ must be slightly broader. 
We also note, that when the distance of the two interfaces of the sample is very close, the integration range of $y$ may include the intersection areas that lead to artifacts. 
Such artifacts cannot be removed by this technique.

The intensity distributions $I(x,y)$ in the dip condition are shown in Figs.~\ref{fig:6-4}(f) and (g), which are the same patterns as those in Figs.~\ref{fig:6-3}(c) and (d).
Integrating the intensity distributions from $y=-50\,\upmu$m to $50\,\upmu$m, we derive the interferograms shown in Figs.~\ref{fig:6-4}(h) and (i).
We can see the dip-shaped, dispersion-insensitive main signals without artifacts in the interferograms, but also the background signals around the main signals as in the case of the peak condition.


We subtract the interferogram in the dip condition from that in the peak condition to extract the main signals from the background signals; we obtain the artifact-free and background-free interferograms (``Peak$-$Dip'' interferograms) shown in Figs.~\ref{fig:6-4}(j) and (k).
For comparison, we also show ``Auto-correlation'' interferograms in Figs.~\ref{fig:6-4}(l) and (m), which are derived from the intensity distributions in the peak condition at $y=0$\,$\upmu$m in Figs.~\ref{fig:6-4}(b) and (g).
The ``Auto-correlation'' interferograms are artifact-free and background-free, but dispersion-sensitive \cite{Pe2007broadband}.
In the ``Peak$-$Dip'' interferograms, the widths of the two peaks are 19.4\,$\upmu$m and 18.4\,$\upmu$m FWHM without dispersion, and 20.9\,$\upmu$m and 21.1\,$\upmu$m FWHM with dispersion.
These values are slightly larger than the theoretical value 15.8\,$\upmu$m for input light with pulse duration 74.5\,fs FWHM, due to the lost bandwidth in SFG; nevertheless, they show substantial dispersion cancellation compared with the ``Auto-correlation'' interferograms.
The distance between the two peaks in the ``Peak$-$Dip'' interferograms is 229.3\,$\upmu$m without dispersion and 228.9\,$\upmu$m with dispersion.
The thickness of the coverglass is obtained by dividing the measured distances 229.3\,$\upmu$m and 228.9\,$\upmu$m by the group index $n\sub{g}(\lambda)=n(\lambda)-\lambda\frac{\dd n}{\dd \lambda}=1.5273$ of BK7 at $\lambda=782$\,nm, which results in 150.1\,$\upmu$m and 149.9\,$\upmu$m, in good agreement with the 150\,$\upmu$m measured with a micrometer.

\section{Experimental demonstrations}\label{sec:experiments-results6}

\begin{figure}
\begin{center}
\includegraphics[width=11cm]{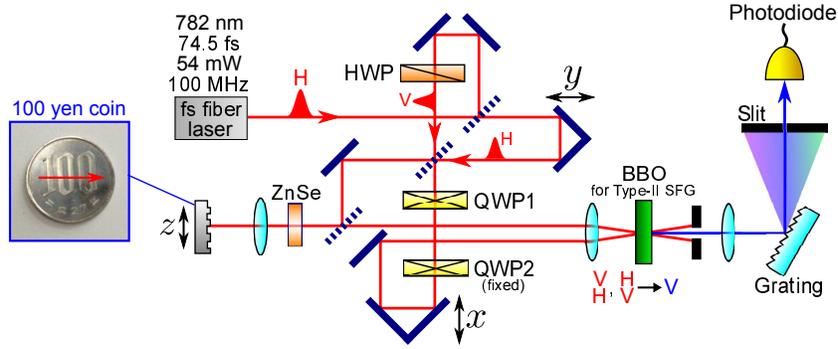}
\caption{
Experimental setup for demonstrating dispersion-canceled and background-free cross-sectional imaging of the surface of a 100-yen coin by TRPI-based OCT with the subtraction method.
The translation $x$, $y$, $z$, and the rotation of QWP1 are controlled by a stage controller.  
}\label{fig:6-6}
\end{center}
\end{figure}

In this section, we experimentally demonstrate dispersion-canceled and background-free cross-sectional imaging of the surface of a coin behind a dispersive layer, as a practical use for TRPI-based OCT with the subtraction method.
The detailed experimental configuration is shown in Fig.~\ref{fig:6-6}.
This system has the same composition as the TRPI-based OCT system with the subtraction method described in Sec.~\ref{sec:removal-artifacts}, except that the coverglass in the sample arm is replaced with a 100-yen coin shown in the inset.
This sample is mounted on a motorized stage so that we can scan the sample transversely by $z$.
QWP1 is mounted on a rotation-motorized stage so that we can switch the peak and dip conditions by program control.
The experimental apparatus expects these motorized stages to be the same as those described in Sec.~\ref{sec:disp-canc-oct}.
We took $x$-direction scans (speed 1\,mm/s, range 400\,$\upmu$m) for every 10\,$\upmu$m in the $y$-direction (from $y=-30$\,$\upmu$m to 30\,$\upmu$m) and summed the scanned data to obtain a single axial scan in the peak and dip conditions.
Subtracting the axial scan in the dip condition from that in the peak condition gives a single dispersion-canceled and background-free axial scan at a single $z$-position. 
We took the axial scans for every 0.1\,mm in $z$-directions to obtain a cross-sectional image of the surface of the coin. 
The acquisition time per image was 75\,minutes, which was limited by the performance of the motorized stages and data acquisition system that we used, but not by the output optical power.
By optimization, the acquisition time can be made much shorter.

\begin{figure}
\begin{center}
\includegraphics[width=31.5pc]{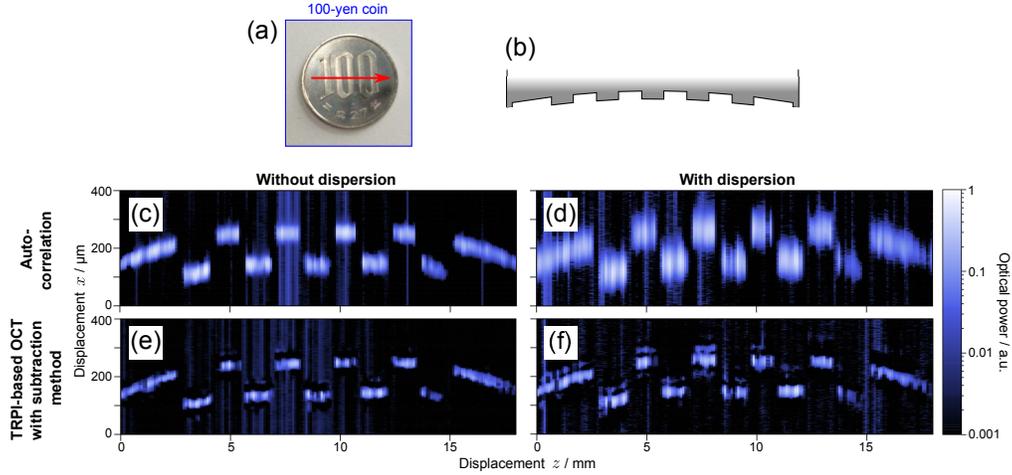}
\caption{(a) 100-yen coin used in the demonstration of dispersion-canceled and background-free cross-sectional OCT.
The red arrow shows the scanned $z$ axis.
(b) Schematic diagram of the cross-sectional structure of the lower half of the coin. 
The surface is slightly curved.
(c)--(f) The measurement results of the cross-sectional OCT images.
The upper [(c), (d)] and lower [(e), (f)] panels show auto-correlation and TRPI-based OCT with the subtraction method, respectively.
The left- [(c), (e)] and right-sided [(d), (f)] panels show the cases without and with dispersion, respectively.
In all the panels, the coin's head is oriented downward. 
The color bar represents the logarithm of the optical power normalized by the maximum value in each panel.
}
\label{fig:6-11}
\end{center}
\end{figure}

The measurement results are displayed in Fig.~\ref{fig:6-11}, where the four panels (c)--(f) show the cases for both auto-correlation and TRPI-based OCT with the subtraction method, without and with dispersion.  
All the images are shown in a log-scale representation.
These images exhibit the corrugated structure of the coin's surface.
Whereas the signal peaks in the case of auto-correlation are significantly broadened, in the presence of dispersion [Figs.~\ref{fig:6-11} (c), (d)], those in the case of TRPI-based OCT with the subtraction method remain essentially unchanged [Figs.~\ref{fig:6-11} (e) and (f)]. 
Note that in this demonstration no artifacts emerge in the measured images, because this measurement is of the  surface topography of one side of the coin.
On the other hand, regardless of the presence of artifacts, background signals remain when integrating the measured data over a limited range of $y$. 
The measurement results in Figs.~\ref{fig:6-11} (e) and (f) show that such background signals are almost removed and that we achieve clear OCT images even in a log-scale representation.
The small remaining background signals seen in Fig.~\ref{fig:6-11} (f) are attributed to the small integral domain of $y$ from $y=-30$\,$\upmu$m to 30\,$\upmu$m and can, therefore, be suppressed by integrating the measured data over a broader domain of $y$. 

\section{Discussion and Conclusion}\label{sec:discussion6}
 
We have proposed TRPI-based OCT with the subtraction method and demonstrated dispersion-canceled, artifact-free, and background-free OCT for axial imaging of a coverglass and cross-sectional imaging of the surface of a coin.
In the demonstration, the measured results exhibit clear OCT images even in a log-scale representation due to the removal of background signals by the subtraction method. 

In the rest of this section, we discuss the relation of this proposal, TRPI-based OCT, to the previously proposed method, CPI-based OCT \cite{Lavoie2009,mazurek2013dispersion}.
As mentioned in the Introduction, both CPI \cite{kaltenbaek2008quantum} and TRPI \cite{PhysRevA.91.013846} are implemented by a time-reversed HOM interferometer, which is a completely classical-optical system and achieves intense output signals.
CPI and TRPI employ pairs of oppositely chirped and transform-limited laser pulses with various time differences, respectively, as their light sources.
%
Compared to CPI-based OCT, TRPI-based OCT has the advantage of being able to be implemented by a simple optical system.
TRPI only has to prepare pairs of transform-limited (i.e., unmodified) laser pulses with various time differences, whereas CPI need to prepare pairs of oppositely chirped laser pulses.
Another advantage of TRPI-based OCT is the removal of the background signals by the subtraction method, which has not been realized in CPI-based OCT.
To apply the subtraction method to CPI-based OCT, we have to label the input chirped and anti-chirped pulses with the polarization degree of freedom, and insert two QWPs into the reference arm to switch the peak and dip conditions, in a manner similar to our experiments.
Then, following the procedure of the subtraction method yields us background-free OCT images even in CPI-based OCT. 
%
The drawback of TRPI-based OCT is that TRPI requires a larger number of measurement processes than CPI, because in TRPI, multiple $x$-direction scans for several values of $y$ are required to obtain a single dispersion-canceled $x$-direction scan.
The number of additional scans is, however, constant (seven, in our experiment in Sec.~\ref{sec:experiments-results6}) with the size of the measured sample; therefore, the number of processes in TRPI-based OCT is considered to be of the same order as that in CPI-based OCT.
Finally, imaging of biological samples, which has been demonstrated by CPI-based OCT \cite{mazurek2013dispersion}, can also be performed by TRPI-based OCT if we use more nonlinear and broadband optical crystals \cite{Pe2007broadband,PhysRevLett.100.183601,Tanaka:12} and more sensitive detectors like single photon counters.
For damage-tolerant samples, we can also employ a laser source with higher power and shorter pulse duration.
TRPI-based OCT with the subtraction method is, in addition to CPI-based OCT, expected to be a useful means for practical OCT.

%




\section*{Acknowledgments}
                                        
This research was supported by JSPS KAKENHI Grant Numbers 25287101 and 15J06614.

\end{document}